\pdfoutput=1
\documentclass[aps,prl,reprint,superscriptaddress]{revtex4-1}
\usepackage{blindtext}
\usepackage{amsmath,amssymb,xspace,graphicx}
\usepackage[utf8]{inputenc} 

\begin{document}
\title{Rolling and ageing in T-ramp soft adhesion}

\author{Giuseppe Boniello}
\affiliation{\'Ecole Normale Sup\'erieure, D\'epartement de Chimie, UMR CNRS-ENS-UPMC 8640 PASTEUR,
24, rue Lhomond,
F-75005 Paris, France}
\author{Christophe Tribet}
\affiliation{\'Ecole Normale Sup\'erieure, D\'epartement de Chimie, UMR CNRS-ENS-UPMC 8640 PASTEUR,
24, rue Lhomond,
F-75005 Paris, France}
\author{Emmanuelle Marie}
\affiliation{\'Ecole Normale Sup\'erieure, D\'epartement de Chimie, UMR CNRS-ENS-UPMC 8640 PASTEUR,
24, rue Lhomond,
F-75005 Paris, France}
\author{Vincent Croquette}
\affiliation{\'Ecole Normale Sup\'erieure, D\'epartement de Physique and D\'epartement de Biologie, Laboratoire de Physique Statistique UMR CNRS-ENS 8550,
24, rue Lhomond,
F-75005 Paris, France}
\author{Dra\v{z}en Zanchi}
\affiliation{\'Ecole Normale Sup\'erieure, D\'epartement de Chimie, UMR CNRS-ENS-UPMC 8640 PASTEUR,
24, rue Lhomond,
F-75005 Paris, France}
\email{zanchi@ens.fr}



\begin{abstract}

Immediately before adsorption to a horizontal substrate, sinking polymer-coated colloids can undergo a complex sequence of landing, jumping, crawling and rolling events.
Using video tracking we studied the soft adhesion to a horizontal flat plate of micron-size colloids coated by a controlled molar fraction $f$ of the polymer PLL-g-PNIPAM which is temperature sensitive. We ramp the temperature from below to above $T_c=32\pm 1^{\circ}$C, at which the PNIPAM polymer undergoes a transition triggering attractive interaction between microparticles and surface.
The adsorption rate, the effective in-plane ($x-y$) diffusion constant and the average residence time distribution over $z$  were extracted from the Brownian motion records during last seconds before immobilisation.
Experimental data are understood within a rate-equations based model that includes ageing effects and includes three populations: the untethered, the rolling and the arrested colloids. We show that pre-adsorption dynamics casts analyze
a characteristic scaling function $\alpha (f)$ proportional to the number of available PNIPAM patches met by soft contact during Brownian rolling.
In particular, the increase of in-plane diffusivity with increasing $f$ is understood: the stickiest particles have the shortest rolling regime prior to arrest, so that their motion is dominated by untethered phase.

\vspace{3mm}

\end{abstract}

\maketitle


During soft adhesion of colloids to a substrate, ends of polymers protruding from engaged surfaces stochastically explore the opposite side, increasing the number of attached contacts, so that the soft contact domain evolves in time \cite{BhattacharyaNature2008,Santore_kalasin2015near}.  Control of these very final events preceding the immobilisation are of interest for research on functional materials, lubrication, cell adhesion etc. \cite{Pine2015,stuart2010emerging,Crocker:2008aa}.
In the present paper we focus on colloids coated by a controlled molar fraction $f$ of T-responsive polymers that switch interactions between colloids and the surface from repulsive to attractive at $T=T_c\approx 32^{\circ}$C. We analyze the details of the colloid Brownian motion occurring just before their adsorption to the surface.
Near-surface Brownian dynamics can be analyzed by direct video 2D or 3D tracking, either by real-time analysis of the diffraction pattern \cite{croquette1,BonielloNature2015}, by total internal reflection microscopy (TIRM) \cite{Prieve199993,Behrens2003,Bevan2005b} or by three-dimensional ratiometric total internal reflection fluorescence microscopy (3-D R-TIRFM) \cite{wall_kihm2004near,choi2007examination}. The mean-square displacement (MSD) reveals how the diffusion is hindered by near-surface effects \cite{FaucheuxPRE1994,Bevan2000,sharma2010high,Chaikin_PRL_2011}. In the 3D tracking the resident time distribution (RTD), corresponding to the $z$-histogram of the trajectory, has been used for studying the particle-surface interaction potential \cite{Bevan2005b}.
If the adsorption (or self-aggregation) is irreversible, as it is in our present case,  the situation is clearly out of-equilibrium process.
In such case, the ageing effects in the pre-adhesion phase were identified
in constant shear flux by Kalasin and Santore
\cite{Santore_kalasin2015near} and for a particle held (and released) by optical tweezers in suspension above a flat surface, by Kumar, Sharma, Ghosh and Bhattacharya\cite{BhattacharyaNature2008,C3SM00097D}.
To bring the system out from equilibrium we use T-switchable attraction between colloids and the surface.
T-responsiveness  of the polymer Poly(N-isopropylacrylamide) (PNIPAM, $T_c= 32\pm 1^{\circ}$C~\cite{cayre2011stimulus}), at varying surface molar fraction $f$, was exploited for analysis of self-association kinetics of  coated microparticles  \cite{Ballauff2011,malinge}.

\begin{figure*}[t]
  \centering
  \includegraphics[width=18cm]{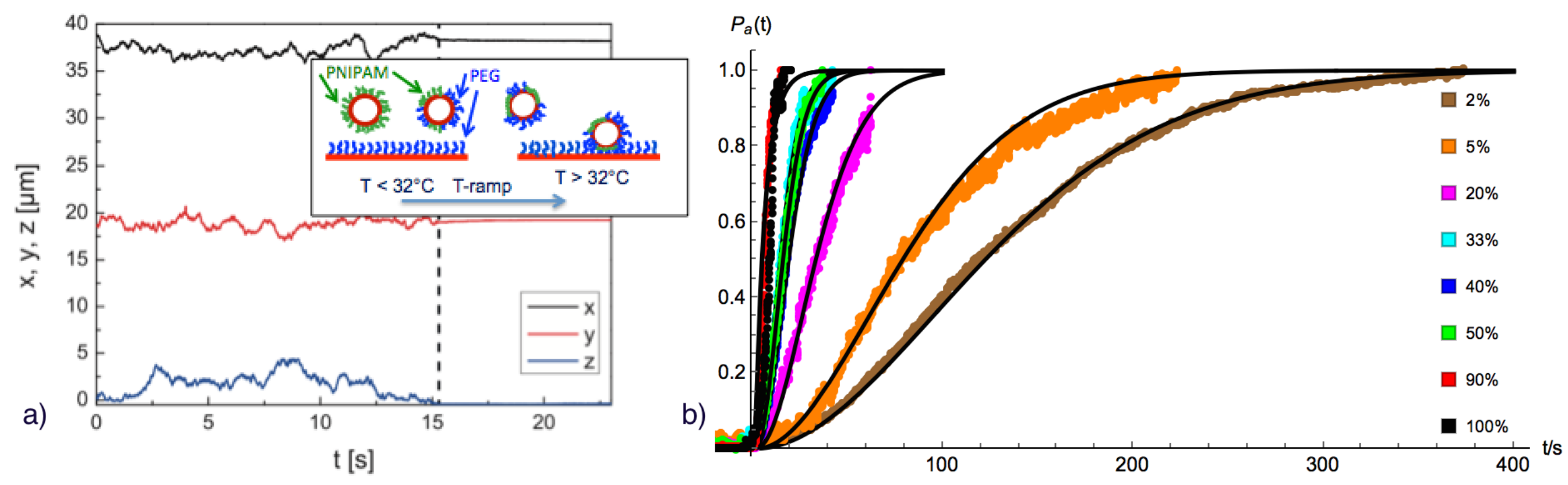}
  \caption{a) Typical 3D tracking record. Bead was captured irreversibly at $t=15.2$ s. Inset: schematic visualisation of temperature ramp experiment. b) Fraction of adsorbed particles as a function of time in a T-ramp of $10^{\circ}$C/min between $26^{\circ}$C and $38^{\circ}$C, for a range of PNIPAM ratios $f$. Solid lines are calculated best fitting adsorption profiles $P_a(t)$.
 }
  \label{fig1prl}
\end{figure*}
\begin{figure}[t]
  \centering
  \includegraphics[width=7cm]{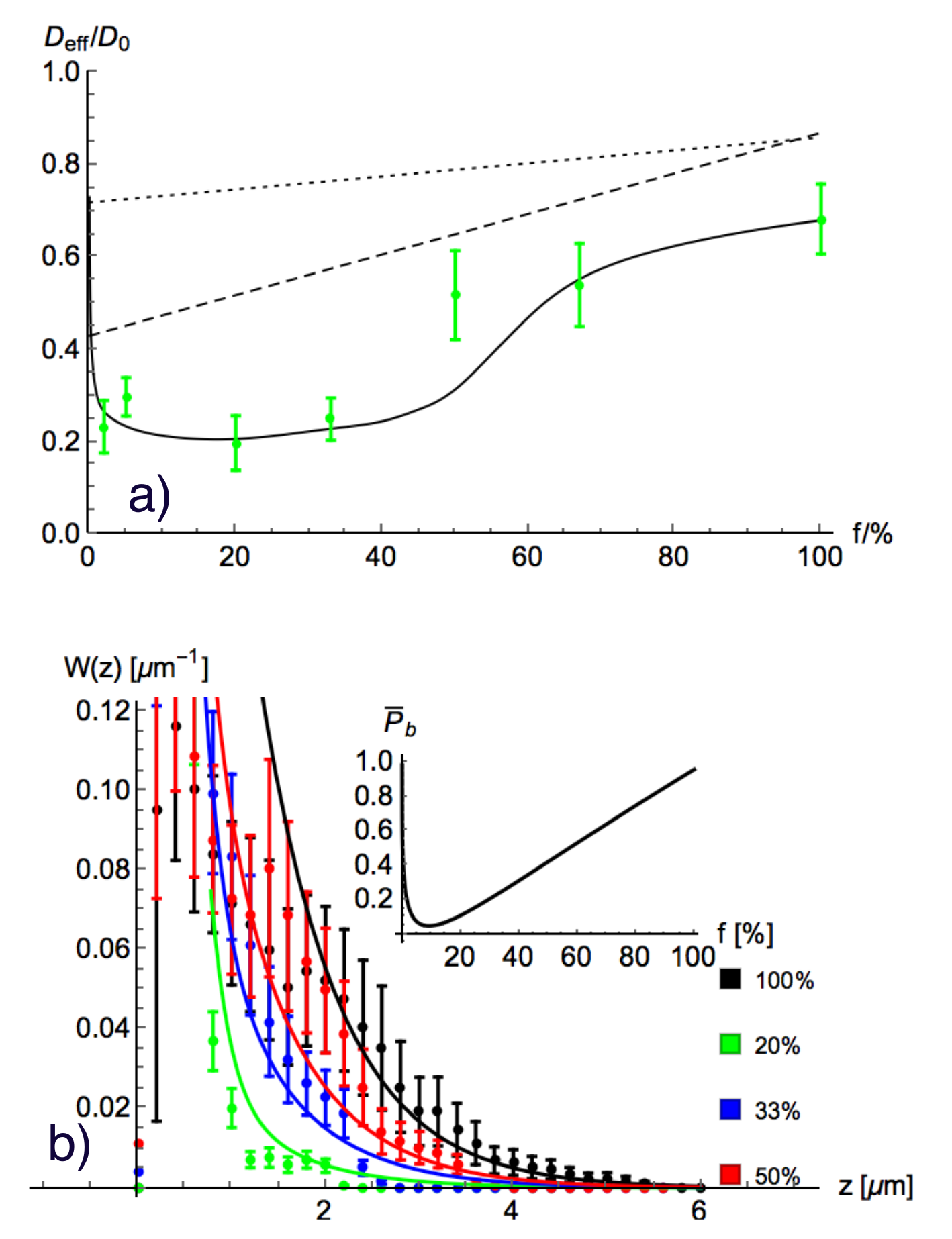}
  \caption{a) Experimental points: effective in-plane diffusion $D_{\mbox{\tiny eff}}(f)$   obtained by fitting the MSDs immediately before stopping. Bold line: result of our theory. Dotted lines: estimation of the near-wall diffusion hindrance by hydrodynamic effects near partially absorbing wall in gravitational potential. Dashed line: same estimation, including Van der Waals potential. b) Residence time distribution $W(z)$ extracted from tracking records $z(t)$ record, compared to calculated profiles. Inset: $P_b(f)$, the calculated average time fraction that particle spend away from the surface during the recording time interval.}
  \label{fig2prl}
\end{figure}

To control the magnitude of T-triggered interactions, we coated silica beads ($0.96\, \mu$m) by molar fraction $f$ of PLL-g-PNIPAM and $(1-f)$ PLL-g-PEG,
(PLL: $M_w = 15-30$ kg/mol; PEG: $M_w = 20$ kg/mol; PNIPAM: $M_w = 7$ kg/mol), see \cite{malinge} for coating conditions.
Aqueous dispersion of beads was deposited on the top of flat glass coverslip, covered by bi-adhesive tape and Mylar film to obtain 52 mm x 5 mm x 50 $\mu$m flow channel. The coverslip was coated with PLL-g-PEG, ensuring a steric repulsion for $T<T_c$.
3D beads motion was observed by slightly defocus microscopy in parallel illumination decorating bead image with interference rings observed with a CMOS camera. Particles were tracked in real time using a PicoTwist apparatus and PicouEye software~\cite{croquette1} allowing sub-pixel resolution ($\sim 5$ nm) at 50 frames/s, see Figure  \ref{fig1prl}a.
Attraction between the beads and the flat plate/surface was triggered by crossing the critical temperature $T_c$, during a T-ramp experiment: the particles suspension was injected in the cell at $26\, ^{\circ}$C, the temperature was increased at $10\,^{\circ}$C/min up to $38\,^{\circ}$C and kept constant until the end of the acquisition.
The fraction of immobilised particles as function of time is shown on Figure \ref{fig1prl}b.
Samples with higher PNIPAM coverage $f$ adsorb faster.
The characteristic sigmoidal shape of the adsorption kinetics indicates that adsorption rate increases gradually, corroborating the ageing nature of the pre-adhesion dynamics of surface engaged particles. If particles are in contact with the surface and rolling (or crawling) most of the time, their in-plane average diffusion $D_{\mbox{\tiny eff}}$  is reduced, as it was observed in experiments with finely thermostated DNA-coated microbeads \cite{Chaikin_PRL_2011}.
In-plane average diffusion $D_{\mbox{\tiny eff}}(f)$ is obtained by fitting the MSDs extracted from in-plane tracking over the last 16 seconds before stopping.  The mean-square displacement (MSD) was extracted from particles $x(t)$ and $y(t)$ tracks. For all values of $f$ MSDs were linear in time over several seconds, indicating that the movement is diffusive.
Figure \ref{fig2prl}a shows $D_{\mbox{\tiny eff}}(f)/D_0$, where $D_0=k_BT/6 \pi \mu R=0.67$ $\mu$m$^2/$ s is the diffusion constant for 1 $\mu$m beads in water far from wall at 38$^\circ$C. As we increase $f$  from $2\%$ to $100\%$, $D_{\mbox{\tiny eff}}/D_0$ increases from $\sim$0.2 to $\sim$0.65.
Naively, one would expect the contrary, that is, stickier are the particles, and slower is their motion. The puzzle is solved by assuming that the surface engaged beads are much slower than the untethered ones, and that the time that beads spend in contact with the surface decreases as $f$ increases. In fact, beads with high coverage stick rapidly after engagement, while these with low $f$ spend most of their time rolling, which lowers their diffusion coefficient considerably.
The RTD over $z$, Fig. \ref{fig2prl}b, obtained by direct 3D particle tracking confirms that beads with high $f$ spend most of their pre-arrest time away from surface.


\begin{figure}[t]
  \centering
  \includegraphics[width=8cm]{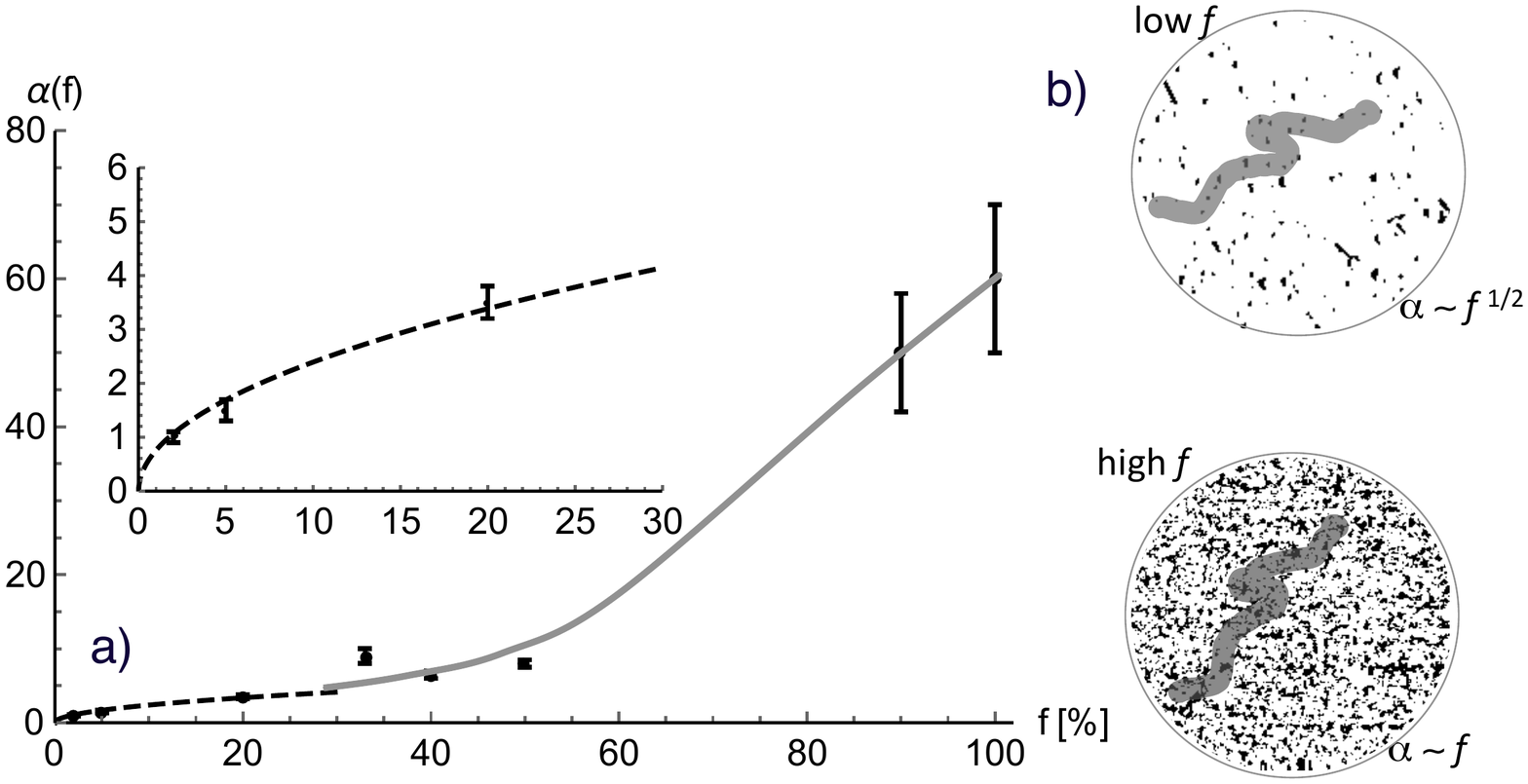}
  \caption{a) Dot symbols: scaling parameter $\alpha(f)$ fitting the experiments. Low $f$ and high $f$ regimes are visible.
  b) Schematic interpretation of the two regimes. Shaded trail is the bead surface portion visited by the contact domain during Brownian rolling. At low $f$ the number of tethers involved is proportional to the linear density of PNIPAM patches along the rolling path, while at high $f$ it crosses over to the surface density.}
  \label{fig3prl}
\end{figure}


According to the theory of Mani, Gopinath and Mahadavan (MGM) \cite{PhysRevLett.108.226104} a typical time scale for soft contact ageing is given by $\tau^* \sim \mu a / n k_BT$, being a result of competition between elastic pulling by $n$ sticky polymer tethers per unit surface and viscous ($\mu$) water draining from the gap between the colloid of radius $a$ and the flat surface.
Moreover, according to \cite{Santore_kalasin2015near,BhattacharyaNature2008,C3SM00097D}, the pre-arrest dynamics of soft adhesion proceeds step-wise, i.e. the 3D approach is well separated from soft contact (2D) rolling phase, indicating that the untethered and the rolling colloids can be seen as separated populations.
Accordingly, to construct the rate equations, we included three populations in the model.
The first is the population of untethered colloids, represented by the probability $P_b(t)$. The second population represents the rolling colloids, $P_r(t)$, while
the third population of colloids is in arrest, $P_a(t)=1-P_b(t)-P_r(t)$. The time zero is chosen to coincide with the onset of the attractive interactions.
The rate equations read  (see Supporting information for details)
\begin{eqnarray}
\dot{P_b} &=&-\kappa P_b+ k_{\mbox{\tiny off}} \, \rho(t/\tau ^*) P_r  \nonumber
 \\
\dot{P_r} &=&\kappa P_b - \left[ k_{\mbox{\tiny off}} \, \rho(t/\tau ^*)
 +k \, g(t/\tau ^*) \right] P_r \; ,
  \label{rateeqn}
 \end{eqnarray}
 where $\kappa=0.17$ s$^{-1}$ is the free sedimentation rate for actual choice of parameters. It corresponds to the fraction of particles absorbed per unit time by a totally absorbing sink \cite{wall_perpendicular_brenner1961slow}.  $k_{\mbox{\tiny off}} \, \rho(t/\tau ^*)$ is the re-dispersion rate of rolling colloids, and  $k \, g(t/\tau ^*)$ is the irreversible stopping rate of rolling colloids.
 $\rho$ and $g$ are the effective "ageing functions":
According to \cite{Santore_kalasin2015near,PhysRevLett.108.226104,C3SM00097D} the "age" can be associated with the number of stuck point contacts within the engaged soft domain.
We chose $\rho(x)=e^{-x}$ and $g(x)=1-\rho(x)$ because we want the re-dispersion rate to decrease and the arrest rate to increase upon ageing.
Initial conditions of the systems are $P_r(0)=P_a(0)=0$ and $P_b(0)=1$.
The central issue of this work is to find out how the parameters of the model, Eqs. \ref{rateeqn}, depend on $f$.
Following MGM, we suppose that the ageing time $\tau ^*$ scales as $n^{-1}$, but with extended meaning of $n$ as the effective surface density of sticky tethers visited by the contact domain during rolling, Fig. \ref{fig3prl}b.
Sticky tethers for us are the dangling polymers that are able to reach the opposite bare surface by overcoming the steric shield. This is possible in the immediate vicinity of the spots containing collapsed PNIPAM. The effective number of available sticky tethers is therefore proportional to the number of PNIPAM spots visited by the contact domain.
For low $f$, we expect that average width of the contact domain is wider than typical distance between PNIPAM spots. The number of tethers involved is proportional to the linear density of PNIPAM patches along the rolling path, i.e. $n\sim f^{1/2}$, while for higher $f$, $n$ becomes proportional to $f$, since the inter-patch distance becomes smaller than the width of the searching trail.
Notice that the present argumentation does not prejudice about details of PNIPAM disposition over the surface, as far as PLL-g-PNIPAM is disposed in a discrete number of spots. In particular, the PLL-g-PNIPAM molecules can be grouped in patches with some size distribution.
In order to confirm that present argumentation makes sense, we assumed that all parameters of Eq.\ref{rateeqn} depend on $f$ over a single, monotonically increasing, scaling function $\alpha(f)$, proportional to the number of available tethers within the searching area. Accordingly,
for the ageing time we pose $\tau^*=\tau^*_0/\alpha(f)$.
Since the stopping rate constant $k$ is supposed to increase with $\alpha(f)$, we put $k=k_0 \alpha(f)$. Detachment rate constant should decrease with increasing number of sticking points: we use $k_{\mbox{\tiny off}}=k_{\mbox{\tiny off0}} / \alpha(f)$.
Calculated $P_a(t)$ using Eqs. \ref{rateeqn} were fitted to adsorption kinetics by adjusting solely the value of $\alpha$ for each $f$, see figure \ref{fig1prl}b. The absolute scale for $\alpha$ being arbitrary, we choose $\alpha(f=2\%)=1$. The fitting parameters are $k_0=k_{\mbox{\tiny off0}}=0.05$ s$^{-1}$ and $\tau^*_0=350$ s for all curves.  The Figure \ref{fig3prl}a shows the best fitting scaling function $\alpha(f)$. It is consistent with $\sim \sqrt{f}$ tendency for low $f$ and is steeper for high $f$, which confirms our picture based on interplay between the discreteness of PNIPAM spots and the finite width of contact domain.

In order to reproduce the measured in-plane diffusion $D_{\mbox{\tiny eff}}$, Fig.\ref{fig2prl}a) we must know when ($t_1$) and for how long $(\Delta t)$ the tracks are recorded because the process is irreversible, so that the average fraction of time that beads spend in "b" and in "r" state is not stationary. We assume that the most probable time $t_1$ is equal $t_{max}(f)$, the time of maximal adsorption rate, corresponding to the maximum slope of kinetics, Fig. \ref{fig1prl}b.
The tracking duration was $\Delta t=16$ s, a compromise regarding the number of particle excursions between "b" and "r" and the lifetime  of  the moving particle starting from $t=0$.
$D_{\mbox{\tiny eff}}$ is calculated as the average of $d_{inst}(t)$ over $\Delta t$ centred at $t_{max}$, which mimics the way in which
experimental $D_{\mbox{\tiny eff}}$ is extracted from the tracking.
Within the present theory, it reads
$D_{\mbox{\tiny eff}}\approx  \frac{1}{\Delta t}\int _{t_{max}-\Delta t/2}^{t_{max}+\Delta t/2} d_{inst}(t)\, dt\;$, where $d_{inst}(t)=[{D_b P_b(t)+D_r P_r(t)}]/[{P_b(t)+P_r(t)}]$,
$D_b$ and $D_r$ are the diffusion constants for untethered and rolling particles respectively
while $P_b(t)$ and $P_r(t)$ are calculated by Eqs. \ref{rateeqn}.
The resulting profile of $D_{\mbox{\tiny eff}}(f)/D_0$, shown on Figure \ref{fig2prl}a) was fitted to experimental data, using values $D_b=0.73 D_0$ and $D_r=0.15 D_0$, while the dependence $\alpha(f)$ (Fig. \ref{fig3prl}a) is the same one that fits the adsorption kinetics, Fig. \ref{fig1prl}b.

In order to fit the RTD to our theory, we associate it to the probability distribution function (PDF), $W(z) = \bar{P_b} W_b(z) + (1 - \bar{P_b}) W_r(z)$
where $\bar{P_b}$ is the average $P_b(t)$ over the observation time of $\Delta t$ centred at $t=t_{max}$,
$\bar{P_b}=\frac{1}{\Delta t}  \int _{t_{max}-\Delta t/2}^{t_{max}+\Delta t/2} \frac{P_b(t)}{P_b(t)+P_r(t)} \, dt$. It is in fact the average fraction of time that particle spends away from the surface during the recording time interval.
We take $W_b(z) \sim e ^{-\tilde{m}gz/k_BT}$ and assume a phenomenological rolling particles distribution $W_r(z) \sim e ^{-a_r z/k_bT}$, with $a_r$ being the apparent weight of rolling particles $a_r \gg \tilde{m}g $.
Calculated distributions $W(z)$ are shown in Figure \ref{fig2prl}b), together with $\bar{P_b}(f)$  in inset.
We see that in the asymptotic part of RTD, corresponding to the untethered particles, is fairly well reproduced by our model, i.e. the untethered population decreases as predicted, with decreasing $f$.

{\em Disussion.} In the light of rate equations-based theory, Eq. \ref{rateeqn}  we understand why the sample with the highest $f$ have also the highest $D_{\mbox{\tiny eff}}$.
At high $f$ a large majority of moving particles are in the suspension far from the surface during tracking, because the rolling regime is very short (i.e. the stopping is faster than the free sedimentation $k\gg \kappa$, so that the beads get arrested as soon as they touch the surface), see inset of Fig. \ref{fig2prl}b. For low $f$ this is not the case any more, the beads spend most of their time engaged in rolling motion, which is much slower.
Eventually, for $f\lesssim 5\%$, the theory predicts a rapid increase in $D_{\mbox{\tiny eff}}(f)/D_0$ with decreasing $f$, as expected, because particles without PNIPAM never stick nor roll on the surface.
The value of $D_b=0.73 D_0$ that fits experimental data should correspond to the equilibrium near-wall hindered diffusion.  The corresponding stationary PDF is $W(z) \propto e ^{-\tilde{m}gz/k_BT}$, where $\tilde{m}$ is the buoyant mass. The diffusion parallel to the surface at distance $z$ is
$D_0 / \phi _{\parallel }(z)$,
where $\phi _{\parallel }(z)$ is reported by \cite{ONeill_67,wall_parallel_goldman1967slow,Book_colloids}. We get
$
D_b=D_{\mbox{\tiny eff}}\mid _{f=0}=D_0\int  \frac{W(z)}{ \phi _{\parallel }(z)} dz=0.72 D_0$, which is fairly close to the best fitting value.  Interestingly, taking into account Van der Waals interaction we get
$D_b=0.44 D_0$, which is too low. This indicates that untethered colloids live in gravitational potential only, since the steric shield prevents them approaching the surface and feeling the VdW forces.
If the calculation is extended to partially absorbing walls implying finite flux solutions of Fokker-Planck equation, we find that $D_{\mbox{\tiny eff}}(f)$ increases linearly, as shown by dotted and dashed lines in Figure
\ref{fig2prl}a), in disagreement with the experimental points, which is an indication that model that ignores the possibility of rolling cannot
explain $D_{\mbox{\tiny eff}}(f)$.
For a similar system, the increase of $D$ upon raising temperature above $T_c$ has been reported in \cite{tu2007brush} in equilibrium conditions, where the phenomenon was attributed to electrostatic effects. This interpretation cannot be applied to our case in which, for high $f$, the particles spend most of their time away from the surface, at distances much higher than the Debye length.

In conclusion, the soft adsorption kinetics depend on $f$ over a single function $\alpha (f)$, which is a measure of the number of discrete sticky patches within the soft contact area during Brownian rolling, extending the theory of Mani, Gopinath and Mahadavan \cite{PhysRevLett.108.226104}. At low/high $f$ the effective inter-patch distance is larger/smaller than the contact diameter. From the point of view of PNIPAM spots (Figure \ref{fig3prl}b) the trail of the rolling contact domain crosses over from 1D to 2D, implying crossover of $\alpha (f)$ from $\sim \sqrt f$ to $\sim f$.
Two most remarkable  effects are (i) a characteristic sigmoidal profile of the adsorption kinetics due to aging, and (ii) a decrease of the in-plane diffusion constant in pre-arrest Brownian dynamics with decreasing $f$ due to reduction of pre-adhesion rolling time.

\vspace{1cm}
\begin{acknowledgments}
We thank Maurizio Nobili for critical reading of the manuscript and Ken Sekimoto for stimulating and helpful discussions. This work was supported by ANR DAPPlePur 13-BS08-0001-01 and program "investissement d'avenir" ANR-11-LABX-0011-01.
\end{acknowledgments}

\newpage

\section{Supporting Information}
We construct the rate equations for three fractions: 1) the untethered fraction represented by the probability $P_b(t)$, 2) the fraction of rolling colloids of age $\tau$, whose probability to be found between the ages $\tau$ and $\tau + d \tau$ is $P_r(t,\tau )\, d\tau$, and
3) the population of colloids in arrest, with the probability
\begin{equation}
P_a(t)=1-P_b(t)-\int _0^t P_r(t,\tau) d\tau \; .
\label{roll1}
\end{equation}
The time zero is chosen to coincide with the onset of the attractive interactions. Consequently, rolling particles cannot be older than $t$.
The rate equation for $P_r(t,\tau)$ writes:
\begin{eqnarray}
 \frac{d P_r(t,\tau)}{dt}&=&K(\tau ) \frac{\partial P_r(t,\tau)}{\partial \tau} + \delta(\tau ) \kappa P_b(t)+\nonumber
 \\
 &-& a(\tau) P_r(t,\tau)-b(\tau) P_r(t,\tau)\; ,
\label{roll2}
\end{eqnarray}
where $K(\tau )$ is the ageing speed, $\delta(\tau )$ is Dirac function, $\kappa$ is the free sinking rate, corresponding to the fraction of particles absorbed per unit time by a totally absorbing bottom. It is determined from the stationary solution of Fokker-Planck equation with $z$-dependent viscous drag\cite{wall_perpendicular_brenner1961slow} and the potential energy given by a sum of Van der Waals and gravitational terms.
$a(\tau)$ is the re-dispersion rate of rolling colloids of age $\tau$ back into bulk, and  $b(\tau)$ is the irreversible stopping rate of rolling colloids of age $\tau$. The first term on the RHS reproduces the homogenous ageing for $K=1$,
 while the second term ensures that any newly sunk particle is a "just born" rolling one ($\tau=0$). It is essential to allow $\tau$ dependence in re-dispersion and arrest rates $a(\tau)$ and $b(\tau)$: it is plausible that older particles, with more sticking contacts, have lower re-dispersion and higher stopping rates.
The rate equation for untethered colloids is
\begin{equation}
\dot{P_b}=-\kappa P_b+\int _0^t  a(\tau) P_r(t,\tau) \, d\tau \; ,
\label{roll3}
\end{equation}
which, together with Eqs. \ref{roll1} and \ref{roll2}, determines completely the evolution of the system.

Since in our experiment the particles are unresolved over $\tau$, starting from Eq.\ref{roll2} a simplified equation is derived, in which all rolling particles are represented by
\begin{equation}
P_r(t)\equiv \int _0^t P_r(t,\tau) d\tau \; .
\label{def}
\end{equation}
We will suppose that the ageing is homogenous ($K(\tau)=1$) and that re-dispersion rate $a(\tau)$ and arrest rate $b(\tau)$ depend on $\tau$ over characteristic aging time scale $\tau ^*$, allowing us to write: $a(\tau) = k_{\mbox{\tiny off}}\Phi (\tau /\tau ^*)$ and $b(\tau) = k \chi (\tau /\tau ^*)$, where $\Phi$ is a monotonically decreasing and $\chi$ monotonically increasing function limited between 0 and 1. In order to obtain Eqs. (1) from the main text, we introduce the effective ageing functions $\rho$ and $g$ as follows:
\begin{equation}
\int _0^t \Phi (\tau /\tau ^*)P_r(t,\tau) d\tau= \rho(t/\tau ^*) P_r(t)
\label{rho}
\end{equation}
 and
\begin{equation}
\int _0^t \chi (\tau /\tau ^*)P_r(t,\tau) d\tau= g(t/\tau ^*) P_r(t) \; .
\label{g}
\end{equation}
These relations are purely formal and do not allow to obtain the functions $\rho$ and $g$ from the original ageing functions $a$ and $b$. In this regard the present theory is merely a phenomenology because the ageing functions are not given explicitly by model parameters. However, the relations \ref{rho} and \ref{g} show that it is always possible to recast the system of rate equations for $P_r(t,\tau)$, to the system for the total number of rolling beads $P_r(t)$, and that the effective ageing functions $\rho$ and $g$ vary at the same aging time scale $\tau ^*$ as the original functions $a$ and $b$.
The equations (1) from the text are readily obtained by integrating over $\tau$ the equation \ref{roll2} and using the definition \ref{def}.


\begin{thebibliography}{26}%
\makeatletter
\providecommand \@ifxundefined [1]{%
 \@ifx{#1\undefined}
}%
\providecommand \@ifnum [1]{%
 \ifnum #1\expandafter \@firstoftwo
 \else \expandafter \@secondoftwo
 \fi
}%
\providecommand \@ifx [1]{%
 \ifx #1\expandafter \@firstoftwo
 \else \expandafter \@secondoftwo
 \fi
}%
\providecommand \natexlab [1]{#1}%
\providecommand \enquote  [1]{``#1''}%
\providecommand \bibnamefont  [1]{#1}%
\providecommand \bibfnamefont [1]{#1}%
\providecommand \citenamefont [1]{#1}%
\providecommand \href@noop [0]{\@secondoftwo}%
\providecommand \href [0]{\begingroup \@sanitize@url \@href}%
\providecommand \@href[1]{\@@startlink{#1}\@@href}%
\providecommand \@@href[1]{\endgroup#1\@@endlink}%
\providecommand \@sanitize@url [0]{\catcode `\\12\catcode `\$12\catcode
  `\&12\catcode `\#12\catcode `\^12\catcode `\_12\catcode `\%12\relax}%
\providecommand \@@startlink[1]{}%
\providecommand \@@endlink[0]{}%
\providecommand \url  [0]{\begingroup\@sanitize@url \@url }%
\providecommand \@url [1]{\endgroup\@href {#1}{\urlprefix }}%
\providecommand \urlprefix  [0]{URL }%
\providecommand \Eprint [0]{\href }%
\providecommand \doibase [0]{http://dx.doi.org/}%
\providecommand \selectlanguage [0]{\@gobble}%
\providecommand \bibinfo  [0]{\@secondoftwo}%
\providecommand \bibfield  [0]{\@secondoftwo}%
\providecommand \translation [1]{[#1]}%
\providecommand \BibitemOpen [0]{}%
\providecommand \bibitemStop [0]{}%
\providecommand \bibitemNoStop [0]{.\EOS\space}%
\providecommand \EOS [0]{\spacefactor3000\relax}%
\providecommand \BibitemShut  [1]{\csname bibitem#1\endcsname}%
\let\auto@bib@innerbib\@empty
\bibitem [{\citenamefont {Sharma}\ \emph {et~al.}(2008)\citenamefont {Sharma},
  \citenamefont {Ghosh},\ and\ \citenamefont
  {Bhattacharya}}]{BhattacharyaNature2008}%
  \BibitemOpen
  \bibfield  {author} {\bibinfo {author} {\bibfnamefont {P.}~\bibnamefont
  {Sharma}}, \bibinfo {author} {\bibfnamefont {S.}~\bibnamefont {Ghosh}}, \
  and\ \bibinfo {author} {\bibfnamefont {S.}~\bibnamefont {Bhattacharya}},\
  }\href {http://dx.doi.org/10.1038/nphys1105} {\bibfield  {journal} {\bibinfo
  {journal} {Nat. Phys.}\ }\textbf {\bibinfo {volume} {4}},\ \bibinfo {pages}
  {960} (\bibinfo {year} {2008})}\BibitemShut {NoStop}%
\bibitem [{\citenamefont {Kalasin}\ and\ \citenamefont
  {Santore}(2015)}]{Santore_kalasin2015near}%
  \BibitemOpen
  \bibfield  {author} {\bibinfo {author} {\bibfnamefont {S.}~\bibnamefont
  {Kalasin}}\ and\ \bibinfo {author} {\bibfnamefont {M.~M.}\ \bibnamefont
  {Santore}},\ }\href {\doibase 10.1021/acs.macromol.5b01977} {\bibfield
  {journal} {\bibinfo  {journal} {Macromolecules}\ }\textbf {\bibinfo {volume}
  {49}},\ \bibinfo {pages} {334} (\bibinfo {year} {2015})}\BibitemShut
  {NoStop}%
\bibitem [{\citenamefont {Wang}\ \emph {et~al.}(2015)\citenamefont {Wang},
  \citenamefont {Wang}, \citenamefont {Zheng}, \citenamefont {Ducrot},
  \citenamefont {Yodh}, \citenamefont {Weck},\ and\ \citenamefont
  {Pine}}]{Pine2015}%
  \BibitemOpen
  \bibfield  {author} {\bibinfo {author} {\bibfnamefont {Y.}~\bibnamefont
  {Wang}}, \bibinfo {author} {\bibfnamefont {Y.}~\bibnamefont {Wang}}, \bibinfo
  {author} {\bibfnamefont {X.}~\bibnamefont {Zheng}}, \bibinfo {author}
  {\bibfnamefont {{\'E}.}~\bibnamefont {Ducrot}}, \bibinfo {author}
  {\bibfnamefont {J.~S.}\ \bibnamefont {Yodh}}, \bibinfo {author}
  {\bibfnamefont {M.}~\bibnamefont {Weck}}, \ and\ \bibinfo {author}
  {\bibfnamefont {D.~J.}\ \bibnamefont {Pine}},\ }\href
  {http://dx.doi.org/10.1038/ncomms8253} {\bibfield  {journal} {\bibinfo
  {journal} {Nat. Commun.}\ }\textbf {\bibinfo {volume} {6}},\ \bibinfo {pages}
  {7253 EP } (\bibinfo {year} {2015})}\BibitemShut {NoStop}%
\bibitem [{\citenamefont {Stuart}\ \emph {et~al.}(2010)\citenamefont {Stuart},
  \citenamefont {Huck}, \citenamefont {Genzer}, \citenamefont {M{\"u}ller},
  \citenamefont {Ober}, \citenamefont {Stamm}, \citenamefont {Sukhorukov},
  \citenamefont {Szleifer}, \citenamefont {Tsukruk}, \citenamefont {Urban}
  \emph {et~al.}}]{stuart2010emerging}%
  \BibitemOpen
  \bibfield  {author} {\bibinfo {author} {\bibfnamefont {M.~A.~C.}\
  \bibnamefont {Stuart}}, \bibinfo {author} {\bibfnamefont {W.~T.}\
  \bibnamefont {Huck}}, \bibinfo {author} {\bibfnamefont {J.}~\bibnamefont
  {Genzer}}, \bibinfo {author} {\bibfnamefont {M.}~\bibnamefont {M{\"u}ller}},
  \bibinfo {author} {\bibfnamefont {C.}~\bibnamefont {Ober}}, \bibinfo {author}
  {\bibfnamefont {M.}~\bibnamefont {Stamm}}, \bibinfo {author} {\bibfnamefont
  {G.~B.}\ \bibnamefont {Sukhorukov}}, \bibinfo {author} {\bibfnamefont
  {I.}~\bibnamefont {Szleifer}}, \bibinfo {author} {\bibfnamefont {V.~V.}\
  \bibnamefont {Tsukruk}}, \bibinfo {author} {\bibfnamefont {M.}~\bibnamefont
  {Urban}},  \emph {et~al.},\ }\href {\doibase 10.1038/nmat2614} {\bibfield
  {journal} {\bibinfo  {journal} {Nat. Mater.}\ }\textbf {\bibinfo {volume}
  {9}},\ \bibinfo {pages} {101} (\bibinfo {year} {2010})}\BibitemShut {NoStop}%
\bibitem [{\citenamefont {Crocker}(2008)}]{Crocker:2008aa}%
  \BibitemOpen
  \bibfield  {author} {\bibinfo {author} {\bibfnamefont {J.~C.}\ \bibnamefont
  {Crocker}},\ }\href {\doibase 10.1038/451528a} {\bibfield  {journal}
  {\bibinfo  {journal} {Nature}\ }\textbf {\bibinfo {volume} {451}},\ \bibinfo
  {pages} {528} (\bibinfo {year} {2008})}\BibitemShut {NoStop}%
\bibitem [{\citenamefont {Gosse}\ and\ \citenamefont
  {Croquette}(2002)}]{croquette1}%
  \BibitemOpen
  \bibfield  {author} {\bibinfo {author} {\bibfnamefont {C.}~\bibnamefont
  {Gosse}}\ and\ \bibinfo {author} {\bibfnamefont {V.}~\bibnamefont
  {Croquette}},\ }\href {\doibase 10.1016/S0006-3495(02)75672-5} {\bibfield
  {journal} {\bibinfo  {journal} {Biophys. J.}\ }\textbf {\bibinfo {volume}
  {82}},\ \bibinfo {pages} {3314} (\bibinfo {year} {2002})}\BibitemShut
  {NoStop}%
\bibitem [{\citenamefont {Boniello}\ \emph {et~al.}(2015)\citenamefont
  {Boniello}, \citenamefont {Blanc}, \citenamefont {Fedorenko}, \citenamefont
  {Medfai}, \citenamefont {Mbarek}, \citenamefont {In}, \citenamefont {Gross},
  \citenamefont {Stocco},\ and\ \citenamefont {Nobili}}]{BonielloNature2015}%
  \BibitemOpen
  \bibfield  {author} {\bibinfo {author} {\bibfnamefont {G.}~\bibnamefont
  {Boniello}}, \bibinfo {author} {\bibfnamefont {C.}~\bibnamefont {Blanc}},
  \bibinfo {author} {\bibfnamefont {D.}~\bibnamefont {Fedorenko}}, \bibinfo
  {author} {\bibfnamefont {M.}~\bibnamefont {Medfai}}, \bibinfo {author}
  {\bibfnamefont {N.~B.}\ \bibnamefont {Mbarek}}, \bibinfo {author}
  {\bibfnamefont {M.}~\bibnamefont {In}}, \bibinfo {author} {\bibfnamefont
  {M.}~\bibnamefont {Gross}}, \bibinfo {author} {\bibfnamefont
  {A.}~\bibnamefont {Stocco}}, \ and\ \bibinfo {author} {\bibfnamefont
  {M.}~\bibnamefont {Nobili}},\ }\href {http://dx.doi.org/10.1038/nmat4348}
  {\bibfield  {journal} {\bibinfo  {journal} {Nat. Mater.}\ }\textbf {\bibinfo
  {volume} {14}},\ \bibinfo {pages} {908} (\bibinfo {year} {2015})}\BibitemShut
  {NoStop}%
\bibitem [{\citenamefont {Prieve}(1999)}]{Prieve199993}%
  \BibitemOpen
  \bibfield  {author} {\bibinfo {author} {\bibfnamefont {D.~C.}\ \bibnamefont
  {Prieve}},\ }\href {\doibase https://doi.org/10.1016/S0001-8686(99)00012-3}
  {\bibfield  {journal} {\bibinfo  {journal} {Adv. Colloid Interface Sci.}\
  }\textbf {\bibinfo {volume} {82}},\ \bibinfo {pages} {93 } (\bibinfo {year}
  {1999})}\BibitemShut {NoStop}%
\bibitem [{\citenamefont {Behrens}\ \emph {et~al.}(2003)\citenamefont
  {Behrens}, \citenamefont {Plewa},\ and\ \citenamefont {Grier}}]{Behrens2003}%
  \BibitemOpen
  \bibfield  {author} {\bibinfo {author} {\bibfnamefont {S.}~\bibnamefont
  {Behrens}}, \bibinfo {author} {\bibfnamefont {J.}~\bibnamefont {Plewa}}, \
  and\ \bibinfo {author} {\bibfnamefont {D.}~\bibnamefont {Grier}},\
  }\href@noop {} {\bibfield  {journal} {\bibinfo  {journal} {Eur. Phys. J. E}\
  }\textbf {\bibinfo {volume} {10}},\ \bibinfo {pages} {115} (\bibinfo {year}
  {2003})}\BibitemShut {NoStop}%
\bibitem [{\citenamefont {Wu}\ and\ \citenamefont {Bevan}(2005)}]{Bevan2005b}%
  \BibitemOpen
  \bibfield  {author} {\bibinfo {author} {\bibfnamefont {H.-J.}\ \bibnamefont
  {Wu}}\ and\ \bibinfo {author} {\bibfnamefont {M.~A.}\ \bibnamefont {Bevan}},\
  }\href {http://dx.doi.org/10.1021/la047892r} {\bibfield  {journal} {\bibinfo
  {journal} {Langmuir}\ }\textbf {\bibinfo {volume} {21}},\ \bibinfo {pages}
  {1244} (\bibinfo {year} {2005})}\BibitemShut {NoStop}%
\bibitem [{\citenamefont {Kihm}\ \emph {et~al.}(2004)\citenamefont {Kihm},
  \citenamefont {Banerjee}, \citenamefont {Choi},\ and\ \citenamefont
  {Takagi}}]{wall_kihm2004near}%
  \BibitemOpen
  \bibfield  {author} {\bibinfo {author} {\bibfnamefont {K.}~\bibnamefont
  {Kihm}}, \bibinfo {author} {\bibfnamefont {A.}~\bibnamefont {Banerjee}},
  \bibinfo {author} {\bibfnamefont {C.}~\bibnamefont {Choi}}, \ and\ \bibinfo
  {author} {\bibfnamefont {T.}~\bibnamefont {Takagi}},\ }\href@noop {}
  {\bibfield  {journal} {\bibinfo  {journal} {Exp. Fluids}\ }\textbf {\bibinfo
  {volume} {37}},\ \bibinfo {pages} {811} (\bibinfo {year} {2004})}\BibitemShut
  {NoStop}%
\bibitem [{\citenamefont {Choi}\ \emph {et~al.}(2007)\citenamefont {Choi},
  \citenamefont {Margraves},\ and\ \citenamefont {Kihm}}]{choi2007examination}%
  \BibitemOpen
  \bibfield  {author} {\bibinfo {author} {\bibfnamefont {C.}~\bibnamefont
  {Choi}}, \bibinfo {author} {\bibfnamefont {C.}~\bibnamefont {Margraves}}, \
  and\ \bibinfo {author} {\bibfnamefont {K.}~\bibnamefont {Kihm}},\ }\href@noop
  {} {\bibfield  {journal} {\bibinfo  {journal} {Phys. Fluids}\ }\textbf
  {\bibinfo {volume} {19}},\ \bibinfo {pages} {103305} (\bibinfo {year}
  {2007})}\BibitemShut {NoStop}%
\bibitem [{\citenamefont {Faucheux}\ and\ \citenamefont
  {Libchaber}(1994)}]{FaucheuxPRE1994}%
  \BibitemOpen
  \bibfield  {author} {\bibinfo {author} {\bibfnamefont {L.~P.}\ \bibnamefont
  {Faucheux}}\ and\ \bibinfo {author} {\bibfnamefont {A.~J.}\ \bibnamefont
  {Libchaber}},\ }\href {\doibase 10.1103/PhysRevE.49.5158} {\bibfield
  {journal} {\bibinfo  {journal} {Phys. Rev. E}\ }\textbf {\bibinfo {volume}
  {49}},\ \bibinfo {pages} {5158} (\bibinfo {year} {1994})}\BibitemShut
  {NoStop}%
\bibitem [{\citenamefont {Bevan}\ and\ \citenamefont
  {Prieve}(2000)}]{Bevan2000}%
  \BibitemOpen
  \bibfield  {author} {\bibinfo {author} {\bibfnamefont {M.~A.}\ \bibnamefont
  {Bevan}}\ and\ \bibinfo {author} {\bibfnamefont {D.~C.}\ \bibnamefont
  {Prieve}},\ }\href@noop {} {\bibfield  {journal} {\bibinfo  {journal} {J.
  Chem. Phys.}\ }\textbf {\bibinfo {volume} {113}},\ \bibinfo {pages} {1228}
  (\bibinfo {year} {2000})}\BibitemShut {NoStop}%
\bibitem [{\citenamefont {Sharma}\ \emph {et~al.}(2010)\citenamefont {Sharma},
  \citenamefont {Ghosh},\ and\ \citenamefont {Bhattacharya}}]{sharma2010high}%
  \BibitemOpen
  \bibfield  {author} {\bibinfo {author} {\bibfnamefont {P.}~\bibnamefont
  {Sharma}}, \bibinfo {author} {\bibfnamefont {S.}~\bibnamefont {Ghosh}}, \
  and\ \bibinfo {author} {\bibfnamefont {S.}~\bibnamefont {Bhattacharya}},\
  }\href@noop {} {\bibfield  {journal} {\bibinfo  {journal} {Appl. Phys.
  Lett.}\ }\textbf {\bibinfo {volume} {97}},\ \bibinfo {pages} {104101}
  (\bibinfo {year} {2010})}\BibitemShut {NoStop}%
\bibitem [{\citenamefont {Xu}\ \emph {et~al.}(2011)\citenamefont {Xu},
  \citenamefont {Feng}, \citenamefont {Sha}, \citenamefont {Seeman},\ and\
  \citenamefont {Chaikin}}]{Chaikin_PRL_2011}%
  \BibitemOpen
  \bibfield  {author} {\bibinfo {author} {\bibfnamefont {Q.}~\bibnamefont
  {Xu}}, \bibinfo {author} {\bibfnamefont {L.}~\bibnamefont {Feng}}, \bibinfo
  {author} {\bibfnamefont {R.}~\bibnamefont {Sha}}, \bibinfo {author}
  {\bibfnamefont {N.}~\bibnamefont {Seeman}}, \ and\ \bibinfo {author}
  {\bibfnamefont {P.}~\bibnamefont {Chaikin}},\ }\href {\doibase
  10.1103/PhysRevLett.106.228102} {\bibfield  {journal} {\bibinfo  {journal}
  {Phys. Rev. Lett.}\ }\textbf {\bibinfo {volume} {106}},\ \bibinfo {pages}
  {228102} (\bibinfo {year} {2011})}\BibitemShut {NoStop}%
\bibitem [{\citenamefont {Kumar}\ \emph {et~al.}(2013)\citenamefont {Kumar},
  \citenamefont {Bhattacharya},\ and\ \citenamefont {Ghosh}}]{C3SM00097D}%
  \BibitemOpen
  \bibfield  {author} {\bibinfo {author} {\bibfnamefont {D.}~\bibnamefont
  {Kumar}}, \bibinfo {author} {\bibfnamefont {S.}~\bibnamefont {Bhattacharya}},
  \ and\ \bibinfo {author} {\bibfnamefont {S.}~\bibnamefont {Ghosh}},\ }\href
  {\doibase 10.1039/C3SM00097D} {\bibfield  {journal} {\bibinfo  {journal}
  {Soft Matter}\ }\textbf {\bibinfo {volume} {9}},\ \bibinfo {pages} {6618}
  (\bibinfo {year} {2013})}\BibitemShut {NoStop}%
\bibitem [{\citenamefont {Cayre}\ \emph {et~al.}(2011)\citenamefont {Cayre},
  \citenamefont {Chagneux},\ and\ \citenamefont {Biggs}}]{cayre2011stimulus}%
  \BibitemOpen
  \bibfield  {author} {\bibinfo {author} {\bibfnamefont {O.~J.}\ \bibnamefont
  {Cayre}}, \bibinfo {author} {\bibfnamefont {N.}~\bibnamefont {Chagneux}}, \
  and\ \bibinfo {author} {\bibfnamefont {S.}~\bibnamefont {Biggs}},\ }\href
  {\doibase 10.1039/C0SM01072C} {\bibfield  {journal} {\bibinfo  {journal}
  {Soft Matter}\ }\textbf {\bibinfo {volume} {7}},\ \bibinfo {pages} {2211}
  (\bibinfo {year} {2011})}\BibitemShut {NoStop}%
\bibitem [{\citenamefont {Zaccone}\ \emph {et~al.}(2011)\citenamefont
  {Zaccone}, \citenamefont {Crassous}, \citenamefont {B\'eri},\ and\
  \citenamefont {Ballauff}}]{Ballauff2011}%
  \BibitemOpen
  \bibfield  {author} {\bibinfo {author} {\bibfnamefont {A.}~\bibnamefont
  {Zaccone}}, \bibinfo {author} {\bibfnamefont {J.~J.}\ \bibnamefont
  {Crassous}}, \bibinfo {author} {\bibfnamefont {B.}~\bibnamefont {B\'eri}}, \
  and\ \bibinfo {author} {\bibfnamefont {M.}~\bibnamefont {Ballauff}},\ }\href
  {\doibase 10.1103/PhysRevLett.107.168303} {\bibfield  {journal} {\bibinfo
  {journal} {Phys. Rev. Lett.}\ }\textbf {\bibinfo {volume} {107}},\ \bibinfo
  {pages} {168303} (\bibinfo {year} {2011})}\BibitemShut {NoStop}%
\bibitem [{\citenamefont {Malinge}\ \emph {et~al.}(2016)\citenamefont
  {Malinge}, \citenamefont {Mousseau}, \citenamefont {Zanchi}, \citenamefont
  {Brun}, \citenamefont {Tribet},\ and\ \citenamefont {Marie}}]{malinge}%
  \BibitemOpen
  \bibfield  {author} {\bibinfo {author} {\bibfnamefont {J.}~\bibnamefont
  {Malinge}}, \bibinfo {author} {\bibfnamefont {F.}~\bibnamefont {Mousseau}},
  \bibinfo {author} {\bibfnamefont {D.}~\bibnamefont {Zanchi}}, \bibinfo
  {author} {\bibfnamefont {G.}~\bibnamefont {Brun}}, \bibinfo {author}
  {\bibfnamefont {C.}~\bibnamefont {Tribet}}, \ and\ \bibinfo {author}
  {\bibfnamefont {E.}~\bibnamefont {Marie}},\ }\href {\doibase
  10.1016/j.jcis.2015.09.016} {\bibfield  {journal} {\bibinfo  {journal} {J.
  Colloid Interface Sci.}\ }\textbf {\bibinfo {volume} {461}},\ \bibinfo
  {pages} {50} (\bibinfo {year} {2016})}\BibitemShut {NoStop}%
\bibitem [{\citenamefont {Mani}\ \emph {et~al.}(2012)\citenamefont {Mani},
  \citenamefont {Gopinath},\ and\ \citenamefont
  {Mahadevan}}]{PhysRevLett.108.226104}%
  \BibitemOpen
  \bibfield  {author} {\bibinfo {author} {\bibfnamefont {M.}~\bibnamefont
  {Mani}}, \bibinfo {author} {\bibfnamefont {A.}~\bibnamefont {Gopinath}}, \
  and\ \bibinfo {author} {\bibfnamefont {L.}~\bibnamefont {Mahadevan}},\ }\href
  {\doibase 10.1103/PhysRevLett.108.226104} {\bibfield  {journal} {\bibinfo
  {journal} {Phys. Rev. Lett.}\ }\textbf {\bibinfo {volume} {108}},\ \bibinfo
  {pages} {226104} (\bibinfo {year} {2012})}\BibitemShut {NoStop}%
\bibitem [{\citenamefont {Brenner}(1961)}]{wall_perpendicular_brenner1961slow}%
  \BibitemOpen
  \bibfield  {author} {\bibinfo {author} {\bibfnamefont {H.}~\bibnamefont
  {Brenner}},\ }\href@noop {} {\bibfield  {journal} {\bibinfo  {journal} {Chem.
  Eng. Sci.}\ }\textbf {\bibinfo {volume} {16}},\ \bibinfo {pages} {242}
  (\bibinfo {year} {1961})}\BibitemShut {NoStop}%
\bibitem [{\citenamefont {O'Neill}(1964)}]{ONeill_67}%
  \BibitemOpen
  \bibfield  {author} {\bibinfo {author} {\bibfnamefont {M.~E.}\ \bibnamefont
  {O'Neill}},\ }\href@noop {} {\bibfield  {journal} {\bibinfo  {journal}
  {Mathematika}\ }\textbf {\bibinfo {volume} {11}},\ \bibinfo {pages} {67}
  (\bibinfo {year} {1964})}\BibitemShut {NoStop}%
\bibitem [{\citenamefont {Goldman}\ \emph {et~al.}(1967)\citenamefont
  {Goldman}, \citenamefont {Cox},\ and\ \citenamefont
  {Brenner}}]{wall_parallel_goldman1967slow}%
  \BibitemOpen
  \bibfield  {author} {\bibinfo {author} {\bibfnamefont {A.~J.}\ \bibnamefont
  {Goldman}}, \bibinfo {author} {\bibfnamefont {R.~G.}\ \bibnamefont {Cox}}, \
  and\ \bibinfo {author} {\bibfnamefont {H.}~\bibnamefont {Brenner}},\
  }\href@noop {} {\bibfield  {journal} {\bibinfo  {journal} {Chem. Eng. Sci.}\
  }\textbf {\bibinfo {volume} {22}},\ \bibinfo {pages} {637} (\bibinfo {year}
  {1967})}\BibitemShut {NoStop}%
\bibitem [{\citenamefont {Russel}\ \emph {et~al.}(1989)\citenamefont {Russel},
  \citenamefont {Saville},\ and\ \citenamefont {Schowalter}}]{Book_colloids}%
  \BibitemOpen
  \bibfield  {author} {\bibinfo {author} {\bibfnamefont {W.~B.}\ \bibnamefont
  {Russel}}, \bibinfo {author} {\bibfnamefont {D.~A.}\ \bibnamefont {Saville}},
  \ and\ \bibinfo {author} {\bibfnamefont {W.~R.}\ \bibnamefont {Schowalter}},\
  }\href@noop {} {\emph {\bibinfo {title} {Colloidal Dispersions}}}\ (\bibinfo
  {publisher} {Cambridge University Press},\ \bibinfo {year}
  {1989})\BibitemShut {NoStop}%
\bibitem [{\citenamefont {Tu}\ \emph {et~al.}(2007)\citenamefont {Tu},
  \citenamefont {Hong}, \citenamefont {Anthony}, \citenamefont {Braun},\ and\
  \citenamefont {Granick}}]{tu2007brush}%
  \BibitemOpen
  \bibfield  {author} {\bibinfo {author} {\bibfnamefont {H.}~\bibnamefont
  {Tu}}, \bibinfo {author} {\bibfnamefont {L.}~\bibnamefont {Hong}}, \bibinfo
  {author} {\bibfnamefont {S.~M.}\ \bibnamefont {Anthony}}, \bibinfo {author}
  {\bibfnamefont {P.~V.}\ \bibnamefont {Braun}}, \ and\ \bibinfo {author}
  {\bibfnamefont {S.}~\bibnamefont {Granick}},\ }\href@noop {} {\bibfield
  {journal} {\bibinfo  {journal} {Langmuir}\ }\textbf {\bibinfo {volume}
  {23}},\ \bibinfo {pages} {2322} (\bibinfo {year} {2007})}\BibitemShut
  {NoStop}%
\end{thebibliography}

%

\end{document}